# Nontrivial temperature behavior of the carrier concentration in the nano-structure "graphene channel - ferroelectric substrate with domain walls"

Anatolii I. Kurchak[1], Anna N. Morozovska[2], Sergei V. Kalinin[3] and Maksym V. Strikha[1,4]

[1] *V.Lashkariov Institute of Semiconductor Physics, National Academy of Sciences of Ukraine, pr. Nauky 41, 03028 Kyiv, Ukraine*

[2] *Institute of Physics of National Academy of Sciences of Ukraine, pr. Nauky 46, 03028 Kyiv, Ukraine*

[3] *The Center for Nanophase Materials Sciences, Oak Ridge National Laboratory, Oak Ridge, TN 37831*

[4] *Taras Shevchenko Kyiv National University, Radiophysical Faculty pr. Akademika Hlushkova 4g, 03022 Kyiv, Ukraine*

## Abstract

This work explores a nontrivial temperature behavior of the carriers concentration, which governs graphene channel conductance in the nano-structure "graphene channel on ferroelectric substrate" that is a basic element for FETs in non-volatile memory units of new generation. We revealed the transition from a single to double antiferroelectric-like hysteresis loop of the concentration voltage dependence that happens with the temperature increase and then exist in a wide temperature range (350 – 500 K). Unexpectedly we revealed the double loops of polarization and concentration can have irregular shape that remains irregular as long as the computation takes place, and the voltage position of the different features (jumps, secondary maxima, etc.) changes from one period to another, leading to the impression of quasi-chaotic behavior. It appeared that these effects originate from the nonlinear screening of ferroelectric polarization by graphene carriers, as well as it is conditioned by the temperature evolution of the domain structure kinetics in ferroelectric substrate. The nonlinearity rules the voltage behavior of polarization screening by graphene 2D-layer and at the same time induces the motion of separated domain walls accompanied by the motion of p-n junction along the graphene channel (2D-analog of Hann effect). Since the domain walls structure, period and kinetics can be controlled by varying the temperature, we concluded that the considered nano-structures based on graphene-on-ferroelectric are promising for the fabrication of new generation of modulators based on the graphene p-n junctions.

# I. Introduction

The discovery of graphene by Novoselov and Geim [1, 2] had attracted great attention of the scientific community to the study of 2D structures, focused researchers on finding ways to minimize electronic devices and accelerated the transition to nanoelectronics [3, 4]. The unique properties of graphene make it possible to talk about its potential applications in nanoelectronics, optoelectronics, nanoplasmonics, and various sensoric devices, as evidenced by many publications, including the use of graphene for gas sensors [5], transparent electrodes for photovoltaic [6], the production of high-speed nonvolatile memory based on field-effect transistors (FET) with a graphene channel [7, 8, 9], and much more.

It is well known that the freestanding graphene cannot be obtained due to the twisting instability. Therefore, all studies associated with this 2D material are necessarily carried out in combination with substrates of various kinds. Among the most common are quartz (or mica) and ferroelectric substrates, at that a pronounced free charge accumulation can take place at the graphene - ferroelectric interface [10, 11]. A large number of studies had focused on layered heterostructures, using graphene, for non-volatile memory elements. However, they mainly addressed the macro characteristics (for the instrument as a whole) inherent in such systems, namely, the appearance of a hysteresis loop of the average concentration of carriers, the width of the memory window, the dependence of the memory window on the switching speed of the gate voltage. However, if we go down one level below and consider the behavior of the carriers when applying an external electric field, their distribution in graphene, the effect of the substrate on the distribution of carriers, then one can obtain more profound knowledge of the processes occurring in the system when applied to an external electric field. These studies are needed to understand and explain the macro-characteristics for such heterostructures.

One of these examples is the existence of a **p-n-junction** in a graphene channel, which was experimentally implemented on a dielectric $SiO_2$ substrate [12, 13], first with the help of a multi-gate system [14]. P-N-junctions in graphene explore the opportunities to observe the physical manifestations of Andreev reflection, Klein tunneling, quantum Hall effect and Veselago lensing [15, 16, 17].

Later on then Hinnefeld et al [18] and Baeumer et al [19] created a p-n-junction in graphene using the ferroelectric substrates Pb(Zr, Ti)$O_3$ and LiNi$O_3$, respectively. In the case of a ferroelectric substrate, the formation of regions with electron and hole concentration of carriers in the channel arises due to the existence of domains with different directions of polarization [20, 21]. Kurchak et al [22, 23] revealed that adsorbed charges dynamics leads to the hysteresis effect on conductivity in the graphene channel on an organic ferroelectric substrate. Kim et al. [24]

propose graphene on ferroelectric for nonvolatile memory and reconfigurable logic-gate operations.

Morozovska et al [25, 26, 27] have shown that the finite-size effects can strongly influence the nonlinear hysteretic dynamics of the stored charge and electro-resistance in the multilayer graphene on ferroelectric with domain stripes of different polarities, which can induce domains with *p*- and *n*-type conductivity, and with p-n-junction potentials at domain walls. Theoretical models for the different types of current regimes (from ballistic to diffusive one) in a single-layer graphene channel at 180°- ferroelectric domain walls have been developed [28, 29] and it was shown that the domain wall contact with the surface creates p-n junction in graphene channel. Recently thermodynamics and kinetics of the conductance of p-n-junctions induced in graphene channel by stripe domains nucleation, motion and reversal in a ferroelectric substrate has been explored using self-consistent approach based on Landau-Ginzburg-Devonshire (**LGD**) theory combined with classical electrostatics, semiconductor theory and quantum statistics for electrotransport calculations [30] at room temperatures.

Taking into account the above-mentioned theoretical results, it should be emphasized the importance and necessity of considering the dynamics of p-n junctions and the distribution of carrier concentrations along the graphene channel for various types of ferroelectrics in a wide temperature range. Such studies can answer the question about the possibility to create the devices based on graphene on ferroelectric heterostructures operating at different temperatures.

Motivated by above problem, this work explores peculiarities of the carriers concentration temperature behavior, which governs the graphene channel conductance in the nano-structure "top gate/dielectric layer/graphene channel/ferroelectric substrate with domain structure" that can be used as a basic element for FETs in non-volatile memory units of new generation. We revealed the nontrivial dependences of the concentration hysteresis on temperature and gate voltage amplitude, which exist in the vicinity of ferroelectric transition temperature and far from it. It appeared that the nontrivial peculiarities originate from the temperature evolution of the domain structure kinetics in ferroelectric substrate.

Geometry of the considered heterostructure is shown in **Fig.1**.

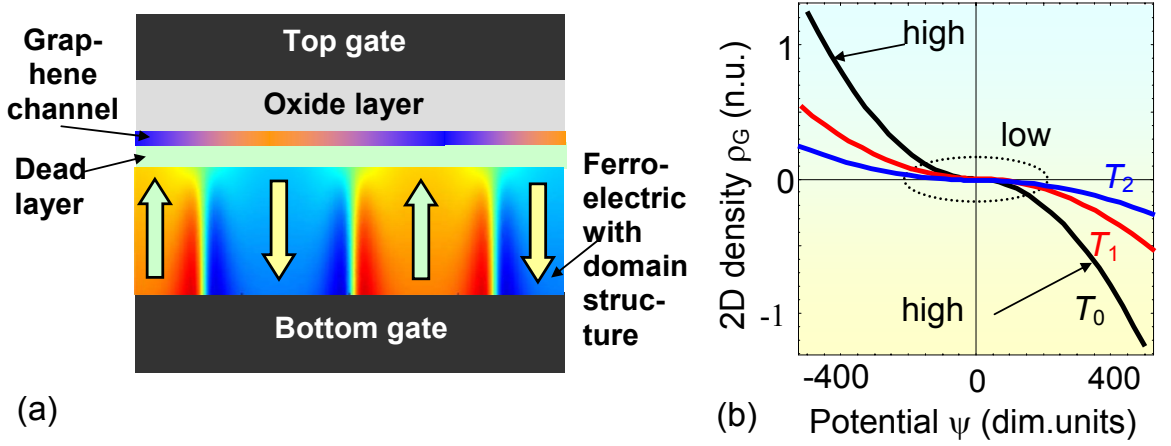

**Figure 1. (a)** Schematics of the nano-structure considered heterostructure "top gate/ dielectric oxide layer / graphene channel / ultra-thin paraelectric dead layer/ ferroelectric substrate with domain walls / bottom gate". (adapted from [30]) **(b)** Dependence of the 2D charge density of graphene on the dimensionless potential ψ calculated at different temperatures $T_0$ (black curve), $T_1 = 1.5T_0$ (red curve) and $T_2 = 2T_0$ (blue curve).

## II. Problem statement

A planar single-layer graphene channel is characterized by the two-dimensional (2D) electron density of states, $g_n(\varepsilon) = g_p(\varepsilon) = 2\varepsilon/(\pi\hbar^2 v_F^2)$ (see e.g. [31, 32]). Corresponding graphene charge density $\sigma_G(\psi) = e(p_{2D}(\psi) - n_{2D}(\psi))$ is the difference of 2D concentrations of electrons in the conduction band [ $n_{2D}(\varphi) = \int_0^\infty d\varepsilon\, g_n(\varepsilon) f(\varepsilon - E_F - e\varphi)$ ] and holes in the valence band [ $p_{2D}(\varphi) = \int_0^\infty d\varepsilon\, g_p(\varepsilon) f(\varepsilon + E_F + e\varphi)$ ], respectively ($E_F$ is a Fermi energy level, these expressions correspond the gapless graphene spectrum). Corresponding Pade-exponential approximation derived in Ref.[30] is

$$\sigma_G(\psi) \approx \frac{2(k_B T)^2 e}{\pi\hbar^2 v_F^2}\left(\frac{1}{\eta(\psi)} - \frac{1}{\eta(-\psi)}\right), \qquad (1)$$

where the functions are $\psi = \dfrac{e\varphi + E_F}{k_B T}$ and $\eta(\psi) = \exp(\psi) + 2\left(\psi^2 + \dfrac{\psi}{2} + \dfrac{2\pi^2}{12-\pi^2}\right)^{-1}$. The density dependence on the dimensionless voltage ψ is shown in **Fig.1(b).**

In the dielectric and oxide dielectric and ultrathin dead layers [33, 34] of thicknesses $h_O$ and $h_{DL}$, respectively, equations of state are $\mathbf{D} = \varepsilon_0 \varepsilon_O \mathbf{E}$ and $\mathbf{D} = \varepsilon_0 \varepsilon_{DL} \mathbf{E}$, $\varepsilon_0$ is a universal dielectric constant. They relate the displacement **D** and electric field **E** in the layers. Since the dead layer is in the paraelectric phase induced by the surface confinement effect, the relative dielectric permittivity of the dead layer $\varepsilon_{DL}$ is rather high $\sim 10^2$ (as it should be for paraelectrics at room temperature [35]) in comparison with unity for a physical gap. The potential $\varphi_{DL}$ satisfies Laplace's equation inside the dead layer.

As a substrate for graphene channel we consider a ferroelectric film of thickness $l$ with the spontaneous polarization component $P_3^f$ directed along its polar axis z. The film contains 180-degree domain wall – surface junctions [see **Fig. 1**]. The dependence of the transverse polarization components $P_1$ and $P_2$ on the field **E** are linear and corresponding relative dielectric permittivities are equal, $\varepsilon_{11}^f = \varepsilon_{22}^f$. Polarization z-component is $P_3(\mathbf{r}, E_3) = P_3^f(\mathbf{r}, E_3) + \varepsilon_0(\varepsilon_{33}^b - 1)E_3$, where a so-called relative "background" permittivity $\varepsilon_{ij}^b \sim$ (4 – 7) is introduced [33, 36]. The distribution of $P_3^f(x, y, z)$ is determined from the time-dependent LGD type Euler-Lagrange equation,

$$\Gamma \frac{\partial P_3^f}{\partial t} + a(T)P_3^f + b(P_3^f)^3 + c(P_3^f)^5 - g\Delta P_3^f = E_3. \tag{2}$$

$\Gamma$ is a Landau-Khalatnikov relaxation coefficient [37], $a(T) = \alpha_T(T - T_C)$, $T$ is the absolute temperature, $T_C$ is a Curie temperature of a bulk ferroelectric, $b$ and $c$ are the coefficients of LGD potential expansion on the polarization powers (also called as linear and nonlinear dielectric stiffness coefficients), $g$ is a gradient coefficient and $\Delta$ stands for a 3D-Laplace operator. The boundary conditions are of the third kind [38],

$$\left(P_3^f - \Lambda_+ \frac{\partial P_3^f}{\partial z}\right)\bigg|_{z=h_D} = 0, \qquad \left(P_3^f + \Lambda_- \frac{\partial P_3^f}{\partial z}\right)\bigg|_{z=h_{DL}+h_F} = 0 \tag{3}$$

The physical range of extrapolation lengths $\Lambda_\pm$ is (0.5 – 2) nm [39]. Quasi-static electric field is defined via electric potential as $E_3 = -\partial \varphi/\partial z$. The potential $\varphi_f$ satisfies Poisson equation inside a ferroelectric film.

For the problem geometry shown in the **Fig. 1** the system of electrostatic equations acquires the form:

$$\Delta \varphi_O = 0, \quad \text{for} \quad -h_O < z < 0, \quad \text{(oxide dielectric layer "O")} \tag{4a}$$

$$\Delta \varphi_{DL} = 0, \quad \text{for} \quad 0 < z < h_{DL}, \quad \text{(dead layer "DL")} \tag{4b}$$

$$\left(\varepsilon_{33}^b \frac{\partial^2}{\partial z^2} + \varepsilon_{11}^f \Delta_\perp\right)\varphi_f = \frac{1}{\varepsilon_0}\frac{\partial P_3^f}{\partial z}, \quad \text{for} \quad h_{DL} < z < h_{DL} + h_F. \quad \text{(ferroelectric "}f\text{")} \tag{4c}$$

3D-Laplace operator is $\Delta$, 2D-Laplace operator on transverse coordinates {x,y} is $\Delta_\perp$. Boundary conditions to Eqs.(4) are as follows: the fixed potential at the top gate, $\varphi_O(x, y, z = -h_O) = U(t)$ and zero potential at the bottom gate $\varphi_f(x,y,z=h_{DL}+h_F)=0$; the continuity of the electric potential at the graphene layer, $\varphi_O(x,y,0)=\varphi_{DL}(x,y,0)$; the equivalence of difference of the electric displacement normal components, and $D_3^{DL}=\varepsilon_0\varepsilon_{DL}E_3$, to the surface charges in graphene $\sigma_G(x,y)$, $D_3^O(x,y,0)-D_3^{DL}(x,y,0)=\sigma_G(x,y)$; and the continuity of the potential and displacement normal components, $D_3^f=\varepsilon_0\varepsilon_{33}^b E_3+P_3^f$ and, at dead layer/ferroelectric interface, $\varphi_d(x,y,h_{DL})=\varphi_f(x,y,h_{DL})$ and $D_3^{DL}(x,y,h_{DL})-D_3^f(x,y,h_{DL})=0$. Electric displacements are $D_3^O=\varepsilon_0\varepsilon_O E_3$, $D_3^{DL}=\varepsilon_0\varepsilon_{DL}E_3$ and $D_3^f=\varepsilon_0\varepsilon_{33}^b E_3+P_3^f$. The gate voltage is periodic with a period $T_g$, $U(t)=U_{\max}\sin(2\pi t/T_g)$. Periodic boundary conditions were applied in x-direction.

### III. Influence of ferroelectric domain structure on the carrier concentration in graphene channel at different temperatures

Below we present results of numerical modeling of the problem (1)-(4). We study numerically the modulation of the carrier concentration, which governs the graphene channel conductance caused by a domain structure moving in a ferroelectric substrate. Parameters used in the calculations are listed in **Table SI**.

Plots in **Fig. 2** show the spatial distribution of polarization in a 75-nm thick ferroelectric substrate calculated at several temperatures within the range (300 – 500)K and gate voltages $U_{\max}$ = 2V **(a)** and 10 V **(b)**. From **Figs.2** the maximal polarization and domain structure contrast decrease and the broadening of the domain walls near the substrate-gap interface increases with temperature increase. From **Fig.2(a)** the domain period decreases with temperature increase. At 300 K the domain structure is relatively contrast in comparison with the faint image at 500 K for $U_{\max}$ = 2 V. The domain period is almost temperature-independent in **Fig.2(b)**, but ferroelectric domain structure disappears at 500 K for $U_{\max}$ = 10 V.

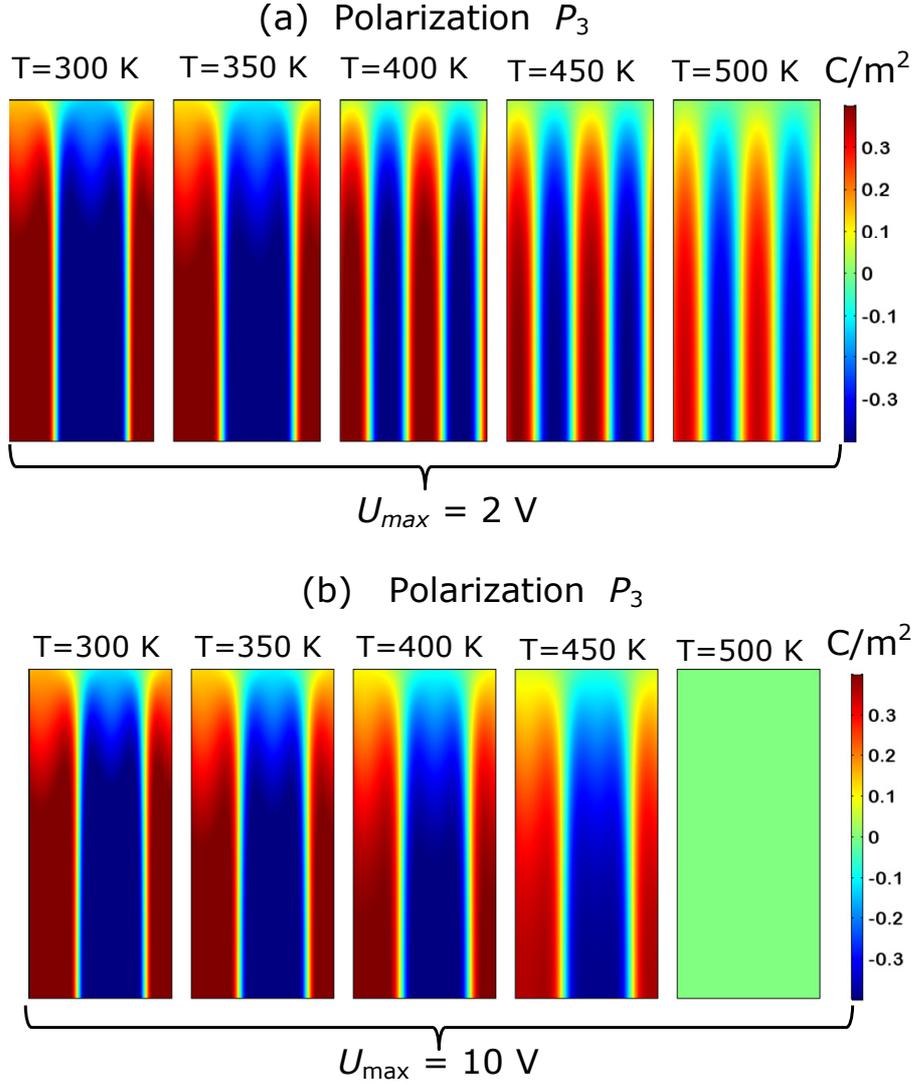

**Figure 2.** Spatial distribution of a ferroelectric polarization component $P_3$ calculated for different temperatures 300, 350, 400, 450 and 500 K (legends at the plots); time $t=T_g=1000$ s. The gate voltage amplitude $U_{max} = 2$ V **(a)** and $U_{max} = 10$ V **(b)**. Parameters are listed in **Table SI.**

The nontrivial question to **Fig.2** is why ferroelectric domain structure disappears at 500 K for $U_{max} = 10$ V, but still exists at $U_{max} = 2$ V? A possible answer is that applied voltage increases the degree of the substrate unipolarity and supports the ferroelectric single-domain state [30]. Note that the transition temperature of the single-domain ferroelectric substrate into a non-polar paraelectric phase decrease under the thickness decrease and is about 500 K for a 70 nm film at $U_{max} = 0$ and chosen material parameters. The critical thickness is much smaller for the poly-domain films [40]. So that increasing $U_{max}$ leads to the film unipolarity and increases its critical thickness at the same time.

The spatial-temporal evolution of the domain structure defines the evolution of the carriers concentration $\Delta n_G(x,t)$ in the graphene channel; at that a separate p-n junction is located above each domain wall [28-30]. To illustrate this, **Figs 3** show x-profiles of the $\Delta n_G(x,t)$ calculated along the graphene channel at several temperatures within the range (300 – 500)K and gate voltages $U_{max}$ = 2V **(a)** and 10 V **(b)**. Positive $\Delta n_G(x,t)$ correspond to holes, and negative to electrons. The amplitude of $\Delta n_G(x,t)$ is maximal for room temperature 300 K, monotonically and strongly decreases with temperature increasing; then it almost [**Fig.3(a)**] or completely [**Fig.3(b)**] vanishes at 500 K. The amplitude decrease is in agreement with the temperature decrease of maximal polarization in the central regions of domains, $P_S \sim \sqrt{T_C - T}$. The modulation amplitude is almost the same at $U_{max}$ = 2V and 10 V.

There are two p-n junctions at (300-350) K located at distance 15 nm; and four p-n junctions at (400-450) K at $U_{max}$ = 2V [**Fig.3(a)**]. The temperature-dependent number of p-n junctions is determined by the corresponding number of domain walls, which doubles with the temperature increase from 300 K to 400 K [see **Fig.3(a)**]. For the reasons described above [see **Fig.2(b)** and comments] the amount of p-n junctions does not change with temperature increase from 300 to 450 K at $U_{max}$ = 10V [see black, red, magenta and blue curves in **Fig.3(b)**]. The charge modulation disappears at 500 K, since the domain structure disappears [see the green line in **Fig.3(b)** and plot for 500 K in **Fig.2(b)**].

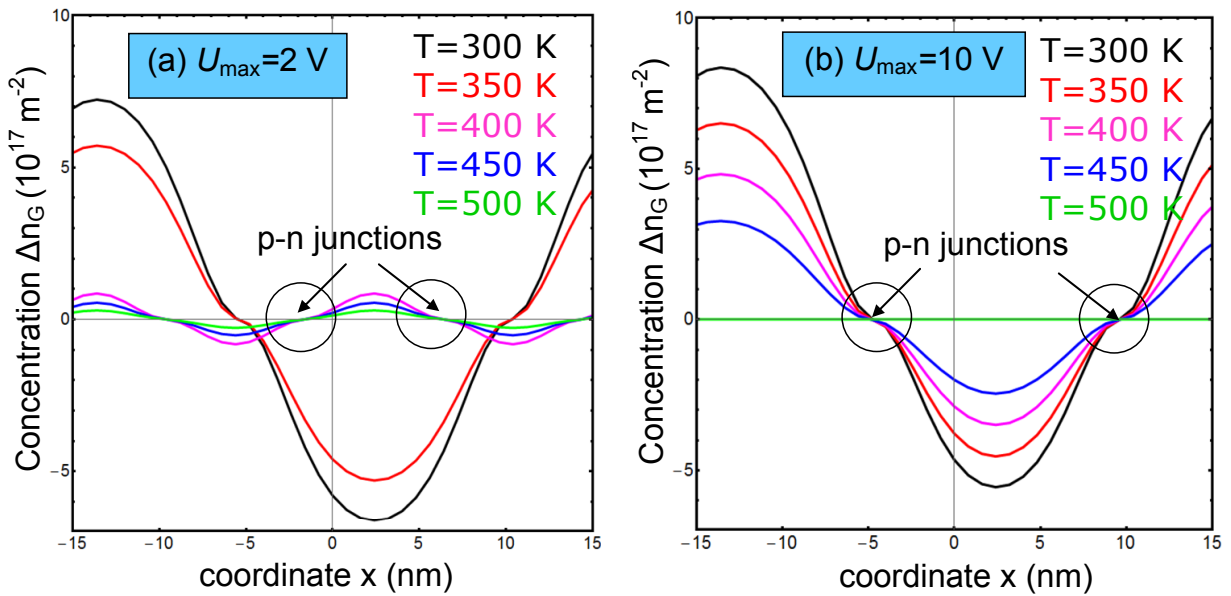

**Figure 3.** 2D-concentration of the carriers $\Delta n_G(x,t)$ in the graphene channel calculated for different temperatures 300, 350, 400, 450 and 500 K (different curves) at time moment $t=T_g=1000$ s. The gate voltage amplitude $U_{max}$ = 2 V **(a)** and $U_{max}$ = 10 V **(b)**. P-N junctions are marked by circles. Parameters are listed in **Table SI.**

Color maps in **Figs 4(a)** and **4(b)** illustrate the temporal evolution of the 2D-concentration of carriers x-profile in graphene channel ($\Delta n_G(x,t)$) for the temperatures 300 K and 400 K, respectively. It is seen that the periodic changes of $\Delta n_G(x,t)$ correlate with the number of domain walls in a ferroelectric substrate that doubles with temperature increasing from 300 K to 400 K. At that the amplitude of $\Delta n_G(x,t)$ decreases in an order of magnitude (from $10^{18}$m$^{-2}$ to $10^{17}$m$^{-2}$) with the temperature increasing from 300 K to 400 K.

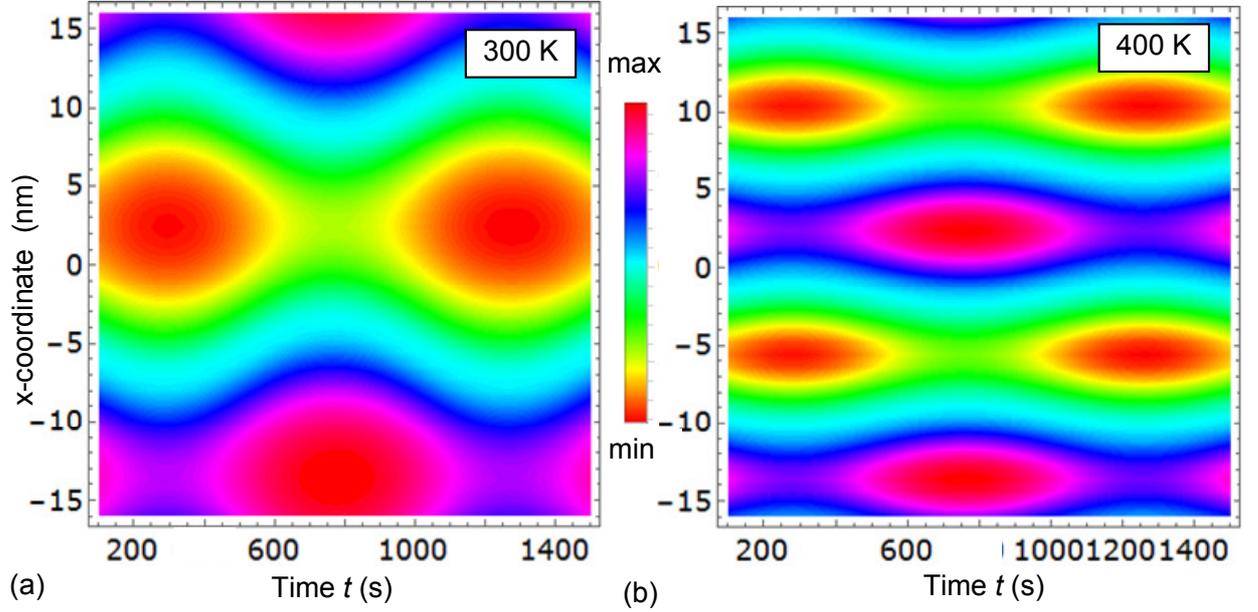

**Figure 4.** Spatial-temporal evolution of the 2D-concentration of carriers in graphene channel calculated for different temperatures 300 K **(a)** and 400 K **(b)**. The gate voltage amplitude $U_{max} = 2$ V. Color scale changes from $\pm 10^{18}$m$^{-2}$ for 300 K to $\pm 10^{17}$ m$^{-2}$ for 400 K. Parameters are listed in **Table SI.**

Color maps in **Figs 5(a)** and **5(b)** show the temperature changes of the 2D-carriers x-profile calculated for two temperature ranges, (300 – 400) K and (300 – 560) K, respectively. It is seen that the contrast of $\Delta n_G(x,t)$ monotonically decreases in an order of magnitude (from $10^{18}$m$^{-2}$ to $10^{17}$m$^{-2}$) with the temperature increasing from 300 K to 560 K. The number of minima and maxima correlate with the number of domain walls in a ferroelectric substrate that doubles with temperature increasing from 300 K to 400 K [see **Fig. 3(a)** and **3(b)**].

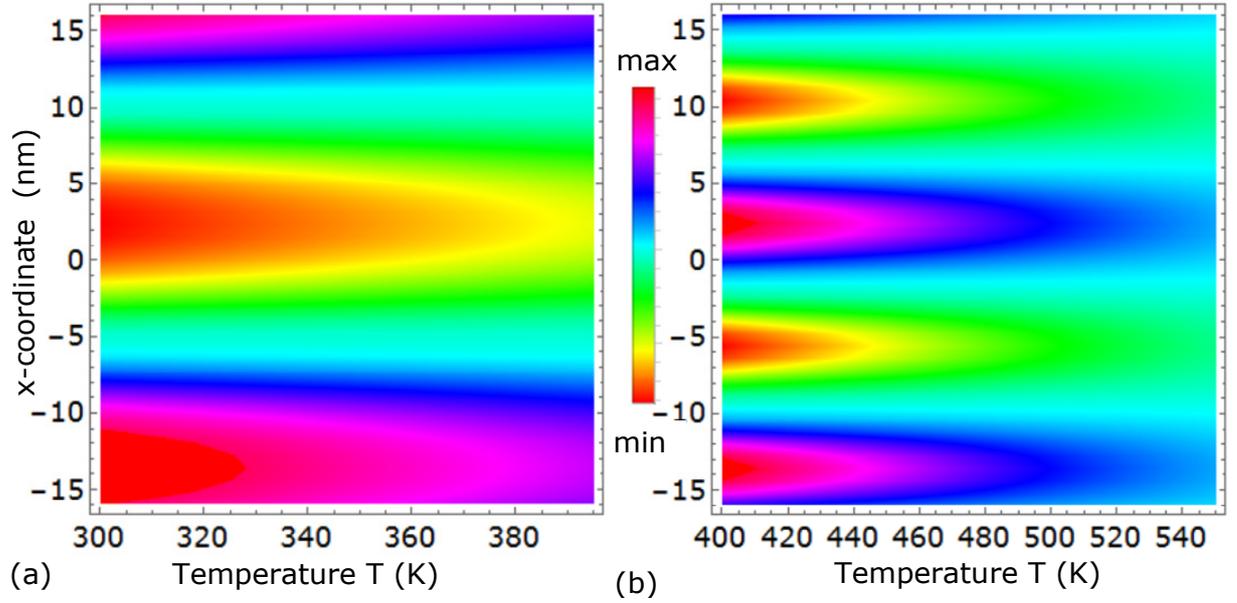

**Figure 5.** Spatial-temperature distribution of the 2D-concentration of carriers in graphene channel calculated for the temperature ranges (300 – 400) K **(a)** and (400 – 560) K **(b)**. The gate voltage amplitude $U_{max}$ = 2 V. Color scale changes from $\pm 10^{18}$ m$^{-2}$ for 300 K to $\pm 10^{17}$ m$^{-2}$ for 400 K. Parameters are listed in **Table SI.**

### IV. Hysteretic response of average carrier concentration in graphene channel

The dependence of the average concentration of carriers in graphene channel on the voltage applied to the gate is hysteretic and temperature-dependent [see the right column in **Fig.6**]. At that the concentration loops are very different from the average polarization loops at small amplitudes of the gate voltage $U_{max} \leq 10$ V [compare **Fig.6(a)** with **6(b)**]. In particular the concentration loop has a pronounced single hysteresis shape with ending characteristic for the conductivity contribution at room temperature and $U_{max} = 10$ V. The loop become lower and slightly thinner at 450 K, then it transforms into a very slim double-shape loop at 500 K (see different loops in **Fig.6(b)** and Ref.[30]). Corresponding polarization loops are very slim in the actual temperature and gate voltage range $U_{max} \leq 10$ V as anticipated for a multi-domain scenario of polarization reversal (see different loops in **Fig.6(a)** and Ref.[30]). Hence the transformation of the concentration loop with the temperature increase is dictated by the domain structure dynamics at $U_{max} \leq 10$ V, rather then by the ferroelectric polarization loop changes, and the relation occurs in a non-proportional way.

Notably the polarization loop shape changes from the slim one and to a pronounced square-like hysteresis with the gate voltage amplitude increase to 15V and temperatures $T \leq 320$ K [see orange and red loops in **Figs. 6(c)**]. At $T \leq 320$ K the temperature behavior of square-like concentration loops is dictated by the almost single-domain polarization reversal at $U_{max} \geq 10$ V, but the relation occurs in a non-proportional way [see orange and red loops in **Figs. 6(d)**].

With the temperature increase above 320 K the double antiferroelectric-like hysteresis loops of polarization and concentration appear, undergoing unusual shape changes (additional flexures and maxima) and exist in a wide temperature range $320 \leq T \leq 500$ K [compare magenta, blue, dark green and light green loops in **Figs. 6(c)** and **6(d)**]. The height and width of the double loops decreases with the temperature increase, and minor loops by-pass is counterclockwise [see magenta, blue, dark green and light green loops in **Figs. 6(e)** and **6(f)**]. These double antiferroelectric-like loops of polarization and concentration disappear at approximately (500 - 550) K indicating the transition of the ferroelectric substrate to a paraelectric phase [see green and black linear curves in **Figs. 6(c)-(f)**].

Note that the wide temperature range of the double loop existence (about 180 K) is at least 10 times more that the range of double hysteresis loops existence in the ferroelectric with the first order phase transition to a paraelectric phase. The physical origin of the double antiferroelectric-like loops can be explained the voltage behavior of the graphene charge density shown in **Fig.1(b)** for different temperatures. The dependence becomes significantly steeper with the temperature decrease. Since the graphene charge is responsible for the polarization screening in a ferroelectric substrate, high charge densities (corresponding to high acting electric potential $|e\varphi| \gg k_B T$ and lower temperatures) can provide an effective screening, and the low ones (corresponding to small potentials $|e\varphi| \leq k_B T$ and higher temperatures) can provide a weak incomplete screening only. At low enough temperatures ($T \leq 320$ K) the substrate (PZT film of thickness 75 nm) is in a deep ferroelectric phase, and relatively weak screening of its depolarization field by graphene sheet is enough to support the ferroelectricity in it. Corresponding hysteresis loop is a single one, has a square-like shape with relatively high remanent polarization and coercive voltage. The situation principally changes with the temperature increase, since the film approaches the paraelectric transition. At that the transition happens at 666 K for a bulk sample, for thin films it happens at much lower temperatures and strongly depends on the film thickness and screening conditions in a self-consistent way [40, 41, 42, 43]. At that the better the screening the higher is the critical thickness of the transition [40, 42, 43]. Thus additional screening by graphene carriers is urgently required to maintain the thin film in a ferroelectric state. As one can see from **Fig.1(b)** the screening increase appears at nonzero potential φ that is in turn proportional to the gate voltage *U*. The critical voltage corresponding to the screening degree enough to suppress the thickness-induced paraelectric transition opens the minor loops of polarization, which in turn induce the concentration loops of anti-hysteretic type.

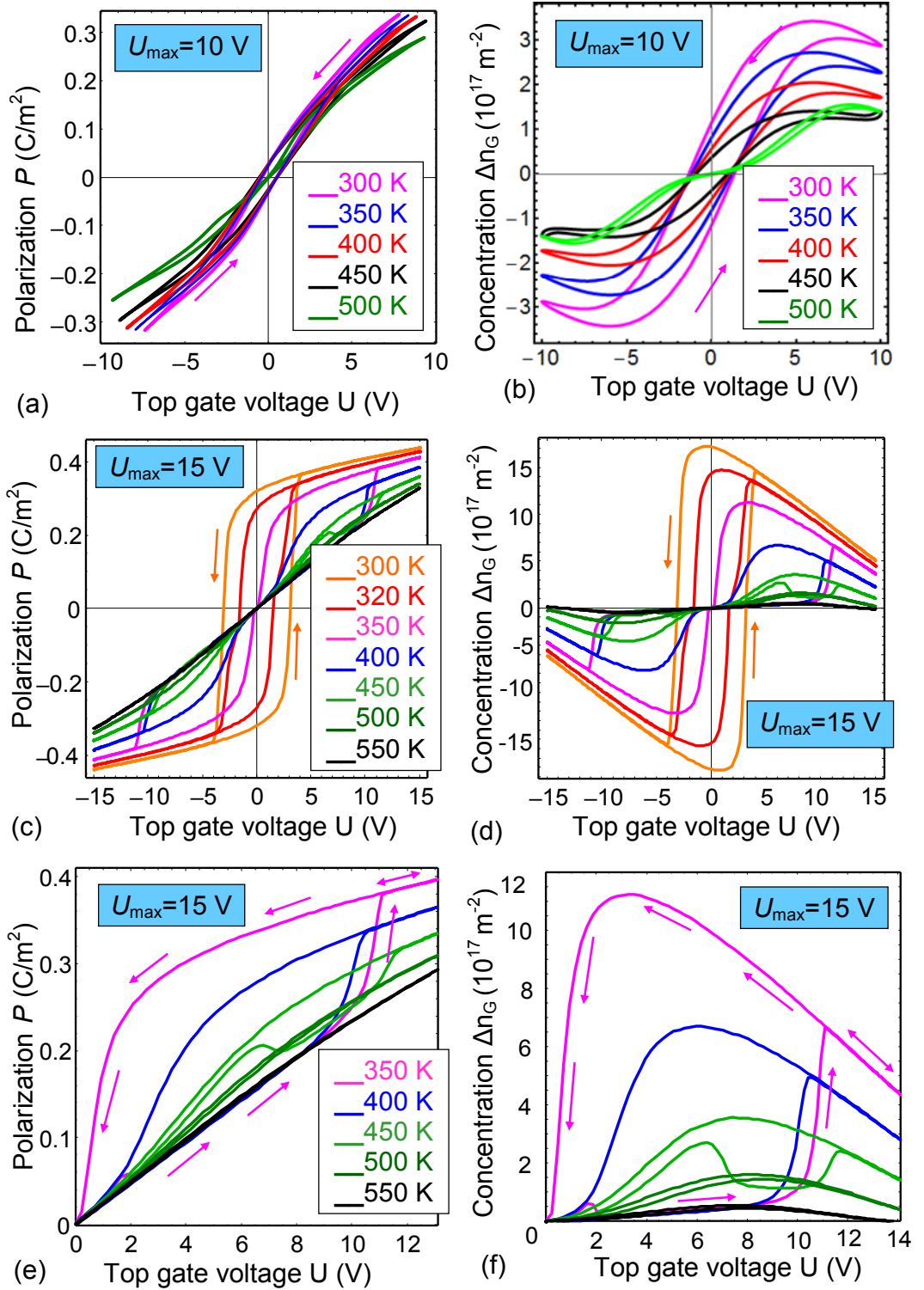

**Figure 6.** Hysteresis loops of the average polarization $P(U)$ of ferroelectric substrate **(a, c, e)** and carriers concentration variation in graphene channel $\Delta n_G(U)$ **(b, d, f)** calculated for different temperatures 300, 320, 350, 400, 450 and 500 K (different curves), $U_{max}$ = 10 V **(a, b)** and $U_{max}$ = 15 V **(c, d, e, f)**; $T_g$=1000 s. Other parameters are listed in **Table SI**. Arrows indicate the counterclockwise

direction of the loops by-pass. Plots **(e)** and **(f)** show positive one-half of the double hysteresis loops calculated at 350, 400, 450, 500 and 550 K(different curves), $U_{max}$ = 15 V.

Single loops of polarization and conductivity (such as shown in **Figs.6**), which correspond either to poly-domain [**Fig.6(a,b)**] or to the monodomain [**Fig.6(c,d)**] scenario of polarization reversal, always have regular shape [see **Figs.7(a)**]. However the double loops, which have regions with poly-domain states at definite temperatures, can be or regular or irregular shape [compare **Figs.7(b)** and **7(c)**].

Single and double hysteresis loops, which have a regular shape (symmetric or anti-symmetric), are reproducible at every period of applied voltage, as anticipated. Quite unexpectedly we revealed the double hysteresis loops of polarization and concentration, which have irregular shape, are "superperiodic" in the sense that they are reproduced only after a great amount of the voltage periods [**Figs.7(c)**]. An example of ten period "superperiodic" polarization and concentration double loops of irregular shape are shown in **Figs.7(d)** and **Figs.7(e)**, respectively. The wording "superperiodicity" in the considered case means that the double loop shape remains irregular as long as the computation takes place, and the voltage position of the different features (jumps, secondary maxima, etc.) changes from one period to another, leading to the impression of quasi-chaotic behavior. We established that the physical origin of the "quasi-chaotic" behavior of the polarization and concentration double loops is the strongly nonlinear voltage behavior of the graphene screening charge [see Eq.(1) and **Fig.1(b)**]. The nonlinearity rules the voltage behavior of polarization screening by graphene 2D-layer and at the same time induces the motion of separated domain walls accompanied by the motion of p-n junction along the channel.

The behavior resembles us 2D-analog of Hann effect in graphene that is almost unexplored. The simplest estimates indicate the possibility of 2D Hunn effect analog, based not on hot carriers redistribution between the sub-bands with different carrier mobility (as in conventional Hunn effect), but on ferroelectric domain wall motion in the substrate. In a conventional Hunn effect one pair of "light" and "heavy" electron domains is moving along the channel and collapses at the drain contact. Immediately afterwards a similar pair occurs at the source electrode. (see, e.g. [44]). This effect in used Hunn diodes for a generation of oscillations with frequency $\nu \approx v_d/L$, where $v_d$ is a velocity of the domain movement, $L$ is a channel length. Generally in GaAs Hunn diodes with $L \sim 10$ μm velocity is $v_d \sim 10^5$ m/s, which leads to frequencies of GHz order. In our case the average velocity of domain wall motion along the channel, $v_{DW} = L/\tau(T) \sim (10^6 - 10^8)$m/s, is determined by ferroelectric polarization relaxation time, $\tau(T) = \Gamma/a(T) \cong \Gamma/\alpha_T(T_C - T)$ [see Eq.(2)], that varies in the range $10^{-13}$s to $10^{-11}$s well below the Curie temperature $T_C$ (as controlled by soft optic phonons). It tends to infinity when $a(T) \to 0$ at temperature $T \to T_C$. Therefore from the inverse time $1/\tau(T)$ we can get the average frequency of THz order and higher for the revealed 2D-analog of Hann effect. However these estimates do not account for the lattice pinning effect on the domain wall motion, that can decrease $v_{DW}$ on several orders of magnitude [45].

To quantify the anomalous temperature behavior of the polarization and graphene carriers concentration loops we calculated the temperature dependences of the switchable remanent polarization $2P_R$ (polarization loop height at zero applied voltage) and concentration $2\Delta n_R$ (concentration loop height at zero applied voltage) and coercive voltage windows, $2U_{CP}$ and $2U_{CN}$, respectively corresponding to the width of the loops [see designations in **Figs.7(a)**]. For double loops we calculated the temperature dependence of the critical voltage of the minor loops opening, $U_{cr}$ [see designations in **Figs.7(b)**].

Calculation results for polarization and concentration loops parameters are shown in **Figs.8** and **9**, respectively**.** As one can from the figures, there are functional proportionally between the parameters temperature behavior, namely $P_R(T) \sim \Delta n_R(T)$ and $U_{CP}(T) \sim U_{CN}(T)$, moreover $U_{CP}(T) = U_{CN}(T)$ for a single loops. However for double loops $U_{CP}(T) < U_{CN}(T)$ on a value about 5-2 V, and the difference is temperature dependent [see **Fig.10**].

Different curves in **Figs.8-9** correspond to different amplitudes of the gate voltage $U_{max} = (2, 5, 8, 10, 15)$ V. Also one can see the pronounced peculiarity related with the transition from the single-domain polarization reversal to a poly-domain one which occurs at $U_{max} = 15$ V and $T \approx 330$ K (see green curves in **Figs.8-9**). The abrupt at black, red, blue and magenta curves corresponds to the transition from the single loop to the double one. The steps at coercive voltages correspond to the domain structure re-building (such as e.g. abrupt changes of its period) taking place due to the spatial confinement of the film in transverse direction. The zigzag-like oscillations of on the polarization, concentration and their coercive fields temperature dependences originated from the quasi-chaotic behavior of the double loops with irregular shape [such as shown in **Figs.7(d)-(e)**]. Namely, the oscillations correspond to the left- and right-voltage asymmetry of the double loops shape.

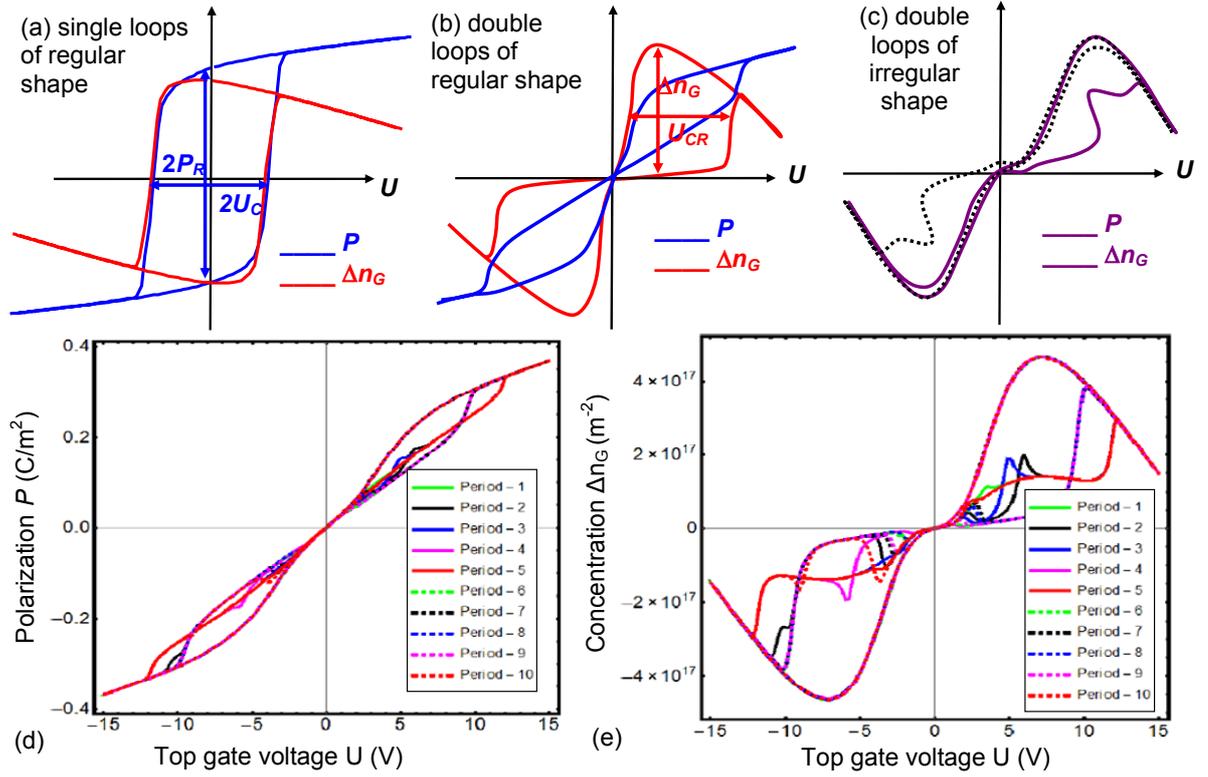

**Figure 7.** Schematics showing the parameters of polarization $P(U)$ **(a)** and concentration $\Delta n(U)$ **(b)** hysteresis loops. Graph **(c)** schematically shows double loops of irregular shape. Multistable quasi-chaotic double loops of polarization **(d)** and concentration **(e)** calculated at T=420K and $U_{max}$ = 15 V. The loops "super-period" consisting of 10 period of the gate voltage.

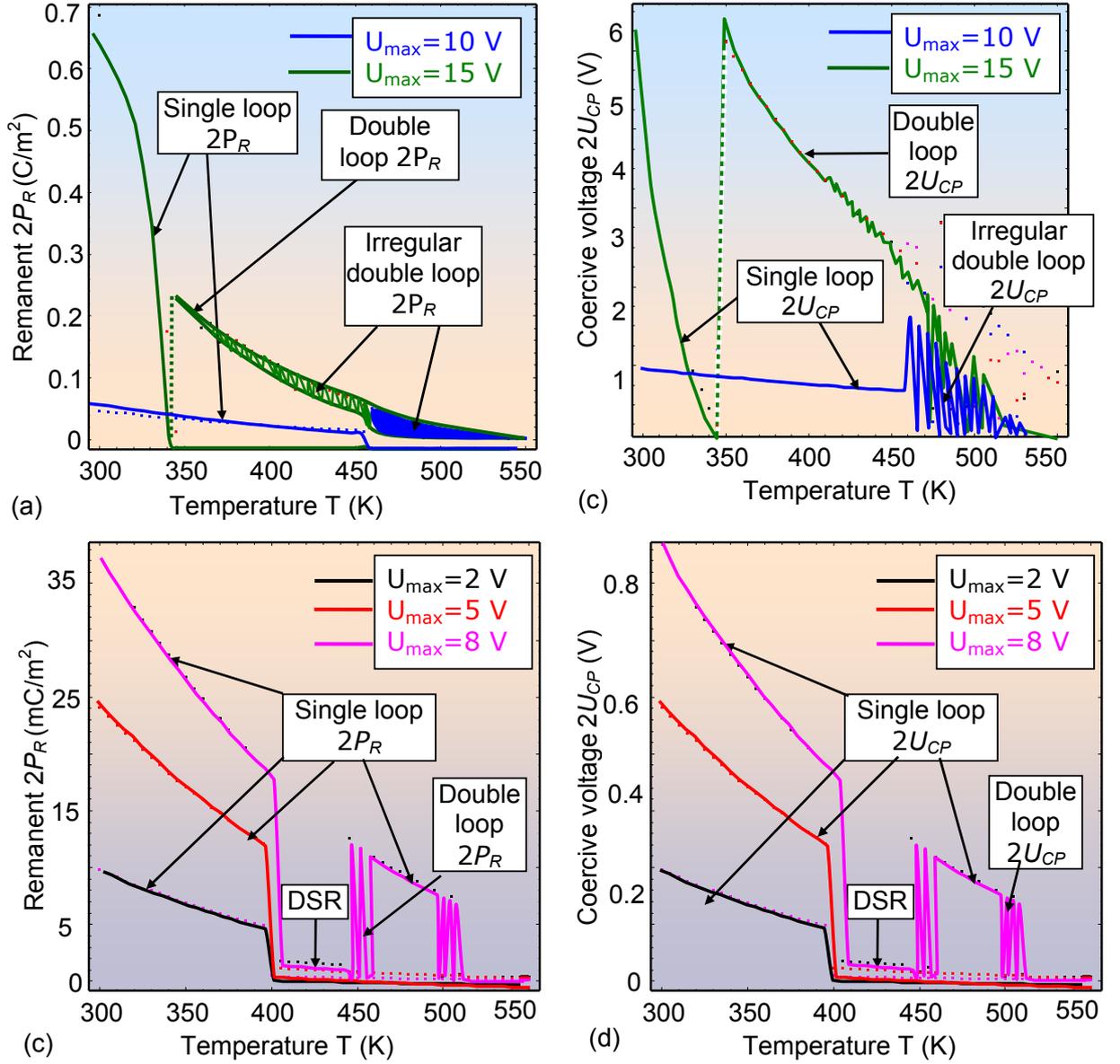

**Figure 8.** Temperature dependences of the remanent average polarization $P_R(U)$ **(a, b),** and coercive voltage of polarization reversal **(c, d)** calculated for the different amplitudes of the gate voltage $U_{max} = (2, 5, 8, 10, 15)$ V and $T_g=1000$ s. Other parameters are listed in **Table SI**. Abbreviation "DSR" indicates the region of domain structure rebuilding. Regularly changing parameters (regular curves) of the single loops and oscillating parameters (randomly oscillating curves) of the double loops with regular or irregular shape are shown.

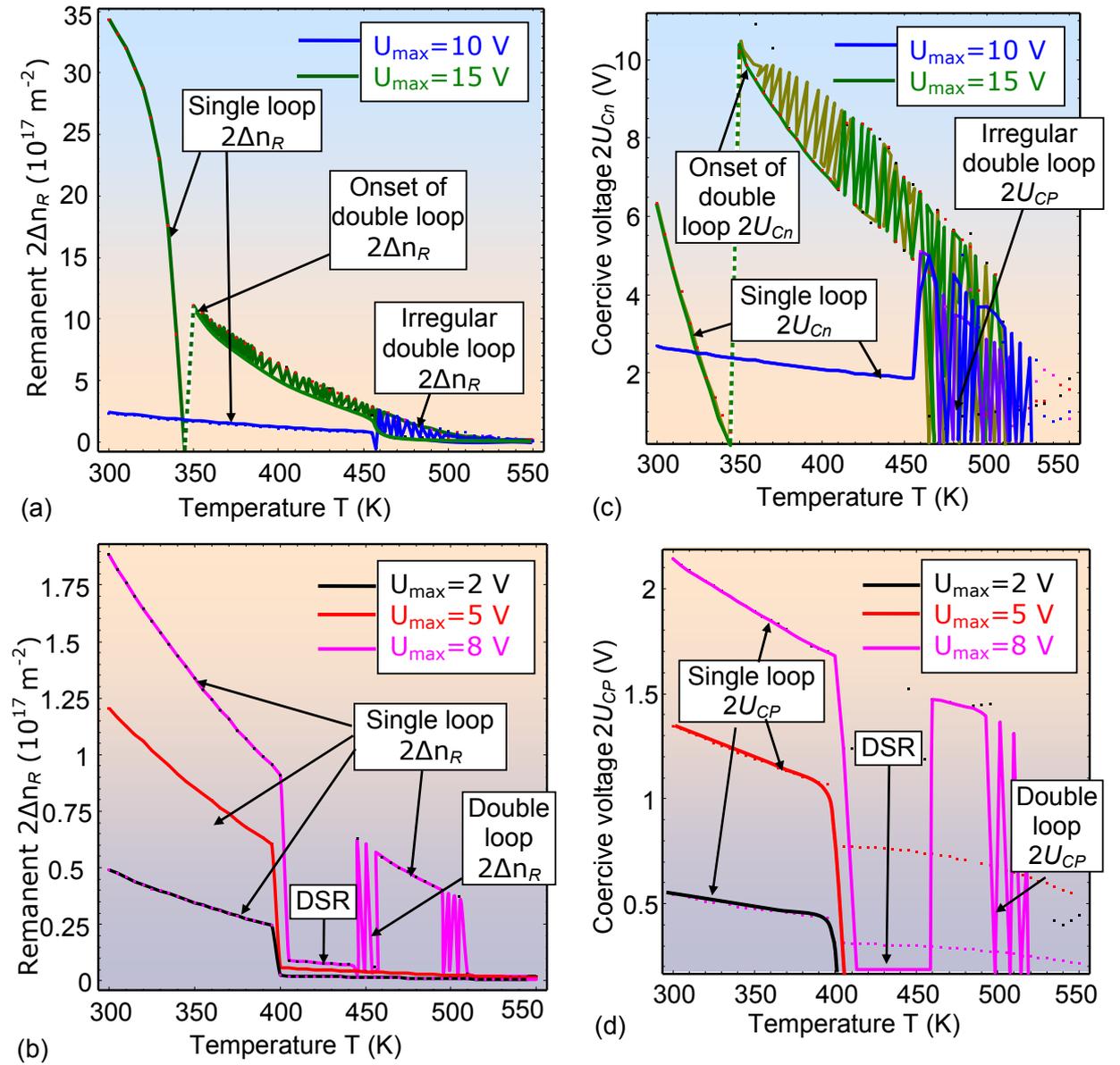

**Figure 9.** Temperature dependences of the remanent concentration of the carriers in graphene channel $\Delta n_R(U)$ **(a, b)** and coercive voltage of concentration reversal **(c,d)** calculated for the different amplitudes of the gate voltage $U_{max} = (2, 5, 8, 10, 15)$ V and $T_g=1000$ s. Other parameters are listed in **Table SI.** Abbreviation "DSR" indicates the region of domain structure rebuilding. Regularly changing parameters (regular curves) of the single loops and oscillating parameters (randomly oscillating curves) of the double loops are shown.

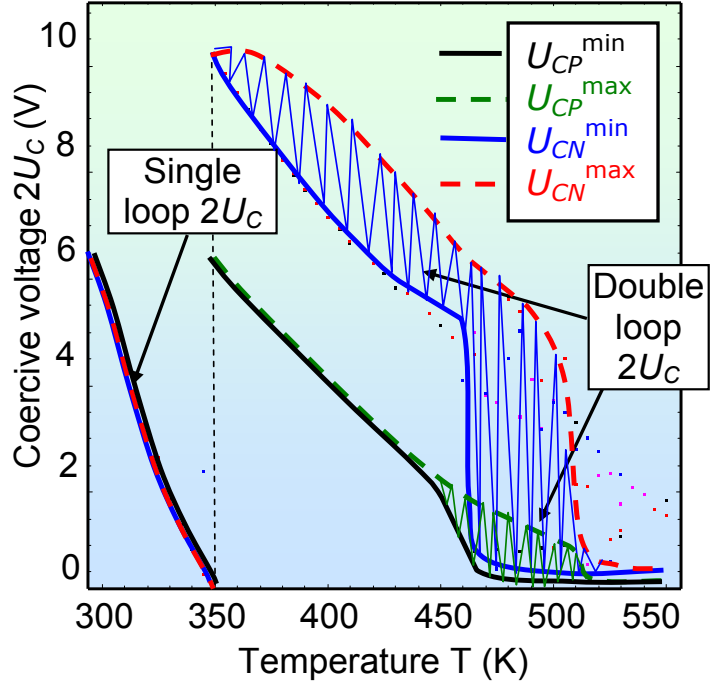

**Figure 10.** Temperature dependences of the coercive voltage of polarization (curves with labels $U_{CP}^{min}$ and $U_{CP}^{max}$) and concentration (curves with labels $U_{CN}^{min}$ and $U_{CN}^{max}$) reversal calculated for the maximal amplitudes of the gate voltage $U_{max}$ = 15 V and $T_g$=1000 s. Other parameters are listed in **Table SI.** The zigzag-like oscillations of on the coercive fields temperature dependences originated from the quasi-chaotic behavior of the double loops with irregular shape

Note, that the problem of the temperature dependence of graphene channel conductance involves also the problem, how the carriers mean free path depends on temperature [3]. However, for the most common case of dominant carriers scattering mechanism in the channel at the ionized impurities in the substrate it can be demonstrated (see e.g. [3, 21]) that the channel conductance follow the general features of temperature dependence, presented by hysteresis loops in **Figs.6-7**, and their parameters, analyzed in **Figs.8-10**. This temperature behavior of the graphene channel conductivity should be taken into account in GFETs operation.

## Conclusions

We studied how the temperature evolution of the domain structure in a ferroelectric substrate and gate voltage amplitude influence on the distribution of the concentration and type of carrier's in a graphene channel. It was found that the average concentration of carriers in graphene channel reveals the behavior similar to ferroelectric (single hysteresis loops) or antiferroelectric-like (double hysteresis loops) depending on the gate voltage amplitude and temperature. At that the wide temperature range of the double loop existence (about 180 K) is at least 10 times more that the range of double hysteresis loops existence in the ferroelectric with the first order phase transition to a paraelectric phase.

The physical origin of the single to double antiferroelectric-like loop transition can be explained the voltage behavior of the graphene charge density shown for different temperatures, at that density becomes significantly steeper with the temperature decrease [**Fig.1(b)**]. The graphene charge is responsible for the polarization screening in a ferroelectric substrate. High charge densities (corresponding to high acting electric potential and lower temperatures) can provide an effective screening, and the low ones (corresponding to small potentials and higher temperatures) can provide a weak incomplete screening only. At room and lower temperatures the substrate (PZT film of thickness 75 nm) is in a deep ferroelectric phase, and relatively weak screening of its depolarization field by graphene sheet is enough to support the ferroelectricity in it. Corresponding hysteresis loop is a single one, has a square-like shape with relatively high remnant polarization and coercive voltage. The situation principally changes with the temperature increase, since the film approaches the paraelectric transition. At that the better the screening the higher is the critical thickness of the transition [41-43]. Thus additional screening by graphene carriers is urgently required to maintain the thin film in a ferroelectric state. As one can see from **Fig.1(b)** the screening increase appears at nonzero potential $\varphi$ that is in turn proportional to the gate voltage $U$. The critical voltage corresponding to the screening degree enough to suppress the thickness-induced paraelectric transition opens the minor loops of polarization, which in turn induce the concentration loops of anti-hysteretic type.

Single and double loops, which have a regular shape, are reproducible at every period of applied voltage, as anticipated. Unexpectedly we revealed the double loops of polarization and concentration, which have irregular shape. They remains irregular as long as the computation takes place, and the voltage position of the different features (jumps, secondary maxima, etc.) changes from one period to another, leading to the impression of quasi-chaotic behavior [**Figs.7(d,e)**]. We established that the physical origin of the "quasi-chaotic" behavior of the polarization and concentration double loops is the strongly nonlinear voltage behavior of the graphene screening charge. The nonlinearity rules the voltage behavior of polarization screening by graphene 2D-layer and at the same time induces the motion of separated domain walls accompanied by the motion of p-n junction along the graphene channel (2D-analog of Hann effect). Since the average velocity of domain wall motion along the channel can be as high as ~$10^7$ m/s well below Curie temperature is determined by ferroelectric polarization relaxation time ~$10^{-12}$s and even smaller, we estimated that the average frequency of operation can be of THz order and higher for the revealed 2D-analog of Hann effect. More detailed analysis of the nonlinear system behavior should be performed within the framework of bifurcation theory [46] that will be the subject of our further studies.

Existence of multi-domain states in a ferroelectric substrate causes inhomogeneous carrier's distribution in graphene channel and lead to formation a set of p-n junctions along channel. Revealed p-n junctions exist in a wide temperature range below the ferroelectric transition temperature; they vanish in the immediate vicinity of the temperature.

Since the domain walls structure, period and kinetics can be controlled by varying the temperature, we concluded that the considered nano-structures based on graphene-on-ferroelectric elements are promising for the fabrication of new generation of modulators and non-volatile memory units based on the graphene p-n junctions.

**Acknowledgements.** The publication contains the results of studies conducted by President's of Ukraine grant for competitive projects (grant number F74/25879) of the State Fund for Fundamental Research (A.I.K., A.N.M.). A.N.M. acknowledges the European Union's Horizon 2020 research and innovation programme under the Marie Skłodowska-Curie grant agreement No 778070. S.V.K. acknowledges the Office of Basic Energy Sciences, U.S. Department of Energy. Part of work was performed at the Center for Nanophase Materials Sciences, which is a DOE Office of Science User Facility.

**Authors' contribution.** A.I.K. wrote the codes, performed numerical calculations and presented their results in graphical form, as well as assisted A.N.M. with the interpretation of numerical results. A.N.M. generated the research idea, stated the problem, and wrote the manuscript draft. S.V.K. and M.V.S. working on the manuscript improvement and discussion.

**Supplementary Materials**

**Table SI.**

| Parameter, constant or value | Numerical value and dimensionality |
|---|---|
| oxide dielectric thickness | $h_O = 8$ nm |
| dielectric (dead) layer thickness | $h_{DL} = 0.4$ nm |
| ferroelectric film thickness | $h_F = 75$ nm |
| graphene channel length | $L$=200 nm (32 nm region is shown in figures) |
| universal dielectric constant | $\varepsilon_0 = 8.85 \times 10^{-12}$ F/m  (e/Vm) |
| permittivity of the dielectric layer | $\varepsilon_{DL} = 100$ (typical range 10 – 300) |
| Ferroelectric permittivity of the ferroelectric film | $\varepsilon_{33}^f = 500$, $\varepsilon_{11}^f = \varepsilon_{22}^f$=780  (Pb(ZrTi)O$_3$-like) |
| Background permittivity of the ferroelectric film | $\varepsilon_{11}^b = \varepsilon_{22}^b = \varepsilon_{33}^b = 4$  (Pb(ZrTi)O$_3$, BaTiO$_3$, or other ferroelectric perovskite) |
| Landau-Ginzburg-Devonshire potential coefficients | $\alpha_T = 2.66\times10^5$ C$^{-2}$·mJ/K, $T_C = 666$ K (PbZr$_x$Ti$_{1-x}$O$_3$, x ≈ 0.5), $P_S^{bulk} = (0.5 - 0.7)$ C/m$^2$ <br> $b = 1.91\times10^8$ J C$^{-4}$·m$^5$, $c = 8.02\times10^8$ J C$^{-6}$·m$^9$ |
| Temperature | $T$ = 300 K; 350 K; 400 K; 450 K; 500 K |
| dielectric anisotropy of ferroelectric film | $\gamma = \sqrt{\varepsilon_{33}^f / \varepsilon_{11}^f} = 0.8$ |
| extrapolation length | $\Lambda_+ = \Lambda_- = \infty$ |
| Plank constant | $\hbar = 1.056\times10^{-34}$ J·s = $6.583\times10^{-16}$ eV·s |
| Fermi velocity of electrons in graphene | $v_F \approx 10^6$ m/s |

# References

[1] K. Novoselov, A. Geim, S. Morozov, D. Jiang, Y. Zhang, S. Dubonos, I. Grigorieva, A. Firsov, "Electric Field Effect in Atomically Thin Carbon Films", Science, **306**, 666 (2004)

[2] A. Geim. "Graphene: status and prospects." Science, **324**, 1530 (2009)

[3] S. Das Sarma, Shaffique Adam, E.H. Hwang, E. Rossi, "Electronic transport in two-dimensional graphene." Rev. Mod. Phys. **83**, 407 (2011)

[4] Gerardo G. Naumis, Salvador Barraza-Lopez, Maurice Oliva-Leyva, and Humberto Terrones. "Electronic and optical properties of strained graphene and other strained 2D materials: a review." *Reports on Progress in Physics* (2017). *arXiv preprint arXiv:1611.08627* (2016).

[5] Schedin, F.; Geim, A. K.; Morozov, S. V.; Hill, E. W.; Blake, P.; Katsnelson, M. I.; Novoselov, K. S. Detection of individual gas molecules adsorbed on graphene. *Nature Materials* **2007**, 6, 652-655.

[6] Guang-Xin, Yi Zheng, Sukang Bae, Chin Yaw Tan, Orhan Kahya, Jing Wu, Byung Hee Hong, Kui Yao and Barbaros Özyilmaz. Graphene – Ferroelectric Hybrid Structure for Flexible Transparent Electrodes. *ACS Nano*, **2012**, *6* (5), 3935–3942

[7] Zhang, X.-W.; Xie, D.; Xu, J.-L.; Zhang, C.; Sun, Y.-L.; Zhao, Y.-F.; Li, X.; Li, X.-M.; Zhu, H.-W.; Chen, H.-M.; Chang, T.-C. Temperature-dependent electrical transport properties in graphene/Pb(Zr$_{0.4}$Ti$_{0.6}$)O$_3$ field effect transistors. *Carbon* **2015**, 93, 384–392.

[8] Rajapitamahuni, A.; Hoffman, J.; Ahn, C. H.; Hong, X. Examining Graphene Field Effect Sensors for Ferroelectric Thin Film Studies. *Nano Letters* **2013**, 13, 4374-4379.

[9] Yusuf, M.H.; Nielsen, B.; Dawber, M.; Du, X. Extrinsic and Intrinsic Charge Trapping at the Graphene/Ferroelectric Interface. *Nano Letters* **2014**, 14, 5437-5444.

[10] Yi Zheng, Guang-Xin Ni, Chee-Tat Toh, Chin-Yaw Tan, Kui Yao, Barbaros Özyilmaz. "Graphene field-effect transistors with ferroelectric gating." Phys. Rev. Lett. **105**, 166602 (2010).

[11] M. Humed Yusuf, B. Nielsen, M. Dawber, X. Du., "Extrinsic and intrinsic charge trapping at the graphene/ferroelectric interface." Nano Lett, **14** (9), 5437 (2014).

[12] V. Cheianov, V. Falko, "Selective transmission of Dirac electrons and ballistic magnetoresistance of n−p junctions in graphene." Phys.Rev.B, **74**, 041403 (2006)

[13] J.R. Williams, L. DiCarlo, C.M. Marcus, "Quantum Hall effect in a gate-controlled pn junction of graphene." Science, **317**, 638 (2007)

[14] Barbaros Ozyilmaz, Pablo Jarillo-Herrero, Dmitri Efetov, Dmitry A. Abanin, Leonid S. Levitov, and Philip Kim. Electronic Transport and Quantum Hall Effect in Bipolar Graphene p−n−p Junctions, Physical Review Letters. 99, 166804(2007).

[15] C.W.Beenakker, "Andreev reflection and Klein tunneling in graphene." Rev.Mod.Phys. **80**, 1337 (2008).

[16] M. I. Katsnelson, K. S. Novoselov, and A. K. Geim, "Chiral tunnelling and the Klein paradox in graphene". Nat. Phys. **2**, 620 (2006).

[17] V. V. Cheianov, V. I. Falko, and B. L. Altshuler, The Focusing of Electron Flow and a Veselago Lens in Graphene *p-n* Junctions. Science **315**, 1252 (2007).

[18] J. H. Hinnefeld, Ruijuan Xu, S. Rogers, Shishir Pandya, Moonsub Shim, L. W. Martin, N. Mason. "Single Gate PN Junctions in Graphene-Ferroelectric Devices." *arXiv preprint arXiv:1506.07138* (2015).

[19] C. Baeumer, D. Saldana-Greco, J. M. P. Martirez, A. M. Rappe, M. Shim, L. W. Martin. "Ferroelectrically driven spatial carrier density modulation in graphene." *Nature communications* **6**, Article number: 6136; doi:10.1038/ncomms7136 (2015).

[20] N.M.Zhang, M.M.Fogler, "Nonlinear screening and ballistic transport in a graphene p − n junction." Phys. Rev. Lett., **100**, 116804 (2008)

[21] Yu. A. Kruglyak, M. V. Strikha. Generalized Landauer – Datta – Lundstrom Model in Application to Transport Phenomena in Graphene. Ukr. J.Phys. Reviews, **10**, 3 (2015)

[22] A.I. Kurchak, M.V. Strikha, "Conductivity of graphene on ferroelectric PVDF-TrFE". Ukr. J. Phys. **59**, 622 – 627 (2014).

[23] Anatolii I. Kurchak, Anna N. Morozovska, and Maksym V. Strikha. Hysteretic phenomena in GFET: general theory and experiment. Journal of Applied Physics, **122**, 044504 (2017)

[24] Woo Young Kim, Hyeon-Don Kim, Teun-Teun Kim, Hyun-Sung Park, Kanghee Lee, Hyun Joo Choi, Seung Hoon Lee, Jaehyeon Son, Namkyoo Park, and Bumki Min. "Graphene-ferroelectric metadevices for nonvolatile memory and reconfigurable logic-gate operations." *Nature communications* **7**, Article number: 10429; doi:10.1038/ncomms10429 (2016).

[25] A. N. Morozovska, M. V. Strikha. "Pyroelectric origin of the carrier density modulation at graphene-ferroelectric interface." J. Appl. Phys. **114**, 014101 (2013).

[26] A. N. Morozovska, E. A. Eliseev, A. V. Ievlev, O. V. Varenyk, A. S. Pusenkova, Ying-Hao Chu, V. Ya. Shur, M. V. Strikha, S. V. Kalinin, "Ferroelectric domain triggers the charge modulation in semiconductors." Journal of Applied Physics, **116**, 066817 (2014).

[27] A. N. Morozovska, A. S. Pusenkova, O.V. Varenyk, S.V. Kalinin, E.A. Eliseev, and M. V. Strikha, "Finite size effects of hysteretic dynamics in multi-layer graphene on ferroelectric". Physical Review B **91**, 235312 (2015).

[28] Anna N. Morozovska, Eugene A. Eliseev, and Maksym V. Strikha. Ballistic conductivity of graphene channel with p-n junction on ferroelectric domain wall. Applied Physics Letters 108, 232902 (2016)

[29] Maksym V. Strikha and Anna N. Morozovska. Limits for the graphene on ferroelectric domain wall p-n-junction rectifier for different regimes of current. J. Appl. Phys. **120**, 214101 (2016)

[30] Anatolii I. Kurchak, Eugene A. Eliseev, Sergei V. Kalinin, Maksym V. Strikha, and Anna N. Morozovska. p−n Junction Dynamics Induced in a Graphene Channel by Ferroelectric-Domain Motion in the Substrate. Phys. Rev. Applied 8, 024027

[31] P. Nemes-Incze, Z. Osváth, K. Kamarás, and L. P. Biró. "Anomalies in thickness measurements of graphene and few layer graphite crystals by tapping mode atomic force microscopy." *Carbon* **46**, no. 11: 1435-1442 (2008).

[32] Elton J.G. Santos, "Electric Field Effects on Graphene Materials." In *Exotic Properties of Carbon Nanomatter*, pp. 383-391. Springer Netherlands, Dordrecht, 2015.

[33] A. K. Tagantsev and G. Gerra. Interface-induced phenomena in polarization response of ferroelectric thin films. J. Appl. Phys. 100, 051607 (2006).

[34] A. K. Tagantsev, M. Landivar, E. Colla, and N. Setter. Identification of passive layer in ferroelectric thin films from their switching parameters. J. Appl. Phys. 78, 2623 (1995).

[35] G. Rupprecht and R.O. Bell, Dielectric constant in paraelectric perovskite. Phys. Rev. 135, A748 (1964).

[36] J. Hlinka and P. Márton, Phenomenological model of a 90° domain wall in BaTiO3-type ferroelectrics. Phys. Rev. B 74, 104104 (2006).

[37] L. D. Landau, and I. M. Khalatnikov. "On the anomalous absorption of sound near a second order phase transition point." Dokl. Akad. Nauk SSSR, vol. 96, pp. 469-472 (1954).

[38] R. Kretschmer and K.Binder. "Surface effects on phase transitions in ferroelectrics and dipolar magnets." Phys. Rev. B **20**, 1065 (1979).

[39] Chun-Lin Jia, Valanoor Nagarajan, Jia-Qing He, Lothar Houben, Tong Zhao, Ramamoorthy Ramesh, Knut Urban & Rainer Waser, "Unit-cell scale mapping of ferroelectricity and tetragonality in epitaxial ultrathin ferroelectric films." Nature Materials, **6**. 64 (2007).

[40] Ivan S. Vorotiahin, Eugene A. Eliseev, Qian Li, Sergei V. Kalinin, Yuri A. Genenko and Anna N. Morozovska. Tuning the Polar States of Ferroelectric Films via Surface Charges and Flexoelectricity. Acta Materialia 137 (15), 85–92 (2017)

[41] Tilley D.R. Finite-size effects on phase transitions in ferroelectrics. Ferroelectic Thin Films. ed. C. Paz de Araujo, J.F.Scott and G.W. Teylor.-Amsterdam: Gordon and Breach, 1996.-P.11-4

[42] E.A. Eliseev, A.N. Morozovska. General approach to the description of the size effect in ferroelectric nanosystems. *The Journal of Materials Science*. 44, № 19, 5149-5160 (2009).

[43] Eugene A. Eliseev, Ivan. S. Vorotiahin, Yevhen M. Fomichov, Maya D. Glinchuk, Sergei V. Kalinin, Yuri A. Genenko, and Anna N. Morozovska. Defect driven flexo-chemical coupling in thin ferroelectric films. (http://arxiv.org/abs/1708.00904)

[44] Simon N. Sze, Kwok N. Ng, Physics of Semiconductor Devices, Wiley (2006)

[45] A.K. Tagantsev, L.E. Cross, and J. Fousek. Domains in ferroic crystals and thin films. New York: Springer, 2010.

[46] Arnold V.I. Theory of catastrophes, 128 p. Moscow: Nauka (1990) In Russian. ISBN 5-02-014271-9